\documentclass[showpacs,aps,prd,longbibliography,nofootinbib,showkeys,superscriptaddress,twocolumn,longbibliography]{revtex4-2}
\usepackage{graphicx}
\usepackage{bm}
\usepackage{multirow}
\usepackage{natbib}
\usepackage{array}
\usepackage{amsmath}
\usepackage{mathrsfs}
\usepackage{amssymb}
\usepackage{float}
\usepackage{xcolor}
\usepackage{enumerate}
\usepackage{color}
\usepackage{slashed}
\usepackage{hyperref}
\usepackage{url}
\usepackage{txfonts}
\usepackage{mathtools}
\usepackage{booktabs}
\usepackage{BOONDOX-cal}
\usepackage{orcidlink}
\usepackage{longtable}
\usepackage{makecell}
\usepackage{cancel}

\allowdisplaybreaks[4]
\hypersetup{hypertex=true,colorlinks=true,linkcolor=blue,anchorcolor=red,citecolor=blue}

\begin{document}
	
	\title{Fully heavy tetraquark states with a diquark-antidiquark configuration}
	
	\author{Xi Xia}
	\affiliation{College of Physics, Chengdu University of Technology, Chengdu 610059, China}
	
	\author{Tao Guo}
	\email[Corresponding author: ]{guot17@tsinghua.org.cn}
	\affiliation{College of Physics, Chengdu University of Technology, Chengdu 610059, China}
	
	\date{\today}

\begin{abstract}

	The study of the fully heavy tetraquark state helps to better understand the characteristics of compact exotic hadrons formed by quarks and gluons. In this work, we systematically investigate the mass spectra and decay channels of fully heavy tetraquark states using a diquark-antidiquark model based on the one-gluon exchange process. The tetraquark states are treated as compact diquark-antidiquark configurations, denoted as $[QQ][\bar{Q}\bar{Q}]$, where $Q = c$ or $b$. By properly accounting for the Pauli principle, basis wave functions incorporating flavor, color, and spin are constructed. The required model parameters are derived from fitting conventional hadron experimental values and lattice QCD calculations. We calculate the mass spectra and decays of the fully heavy tetraquarks in the lower $S$ and $P$ waves. Our results do not support interpreting $X(6600)$, $X(6900)$, and $X(7200)$ as the $1S$-wave fully charmed tetraquark states.  Nevertheless, our analysis suggests that $X(6200)$ can be a good candidate for the $1S$-wave fully charmed tetraquark state with quantum numbers $2^{++}$. Furthermore, we also identify several narrow fully heavy tetraquark states that may be observable in future experiments.

\end{abstract}

\maketitle

\section{Introduction}

Quantum chromodynamics (QCD) allows for a variety of exotic hadronic structures beyond conventional mesons and baryons, such as multiquark states, glueballs, and hybrid states \cite{Godfrey:1985xj,Klempt:2007cp,Lebed:2016hpi}. 
Among these, tetraquark states have become a major focus in hadron physics. 
Since the Belle Collaboration's discovery of the $X(3872)$ resonance in 2003 \cite{Belle:2003nnu}, a growing number of exotic hadrons have been observed experimentally \cite{ParticleDataGroup:2024cfk}. 
These discoveries not only confirm the existence of exotic hadronic states but also inspire further theoretical and experimental investigations into their properties.
Among the tetraquark states, fully heavy tetraquarks represent a particularly intriguing class. 
In such systems, light-meson exchange is forbidden, and the interaction is dominated by short-range color forces mediated by gluon exchange, suggesting a compact multiquark configuration.

In recent years, several candidates for fully charmed tetraquark states have been observed experimentally.
In 2020, the LHCb Collaboration observed a broad structure above twice the $J/\psi$ mass threshold ranging from 6.2 to 6.8 GeV and a narrower structure at about 6.9 GeV, called $X(6900)$ \cite{LHCb:2020bwg}. 
Subsequently, the CMS and ATLAS Collaborations also researched the di-$J/\psi$ channel and both of them confirmed the existence of the $X(6900)$ \cite{CMS:2023owd,ATLAS:2023bft}. 
Notably, the CMS Collaboration also reported two new structures, $X(6600)$ and $X(7200)$, in the di-$J/\psi$ invariant mass spectrum \cite{CMS:2023owd}. 
The ATLAS Collaboration also observed the new structures, $X(6400)$ and $X(6600)$, in the di-$J/\psi$ channel and $X(7200)$ in the $J/\psi$-$\psi(2S)$ invariant mass spectrum \cite{ATLAS:2023bft}. 
Furthermore, analysis of LHCb data generating di-$J/\Psi$ in proton-proton collisions within a coupled-channel framework based on double-vector-charmonium channels indicates the existence of $X(6200)$ \cite{Song:2024ykq}.
Very recently, the CMS Collaboration determined the quantum numbers of these three fully charmed tetraquark states $X(6600)$, $X(6900)$, and $X(7100)$ to be $J^{PC}=2^{++}$ \cite{CMS:2025fpt}.
These experiments have motivated a large number of theoretical studies on fully charmed tetraquark states \cite{Bedolla:2019zwg,Zhang:2020xtb,Wang:2022xja,Dong:2022sef,Faustov:2022mvs,Dong:2020nwy,Liang:2021fzr,Yu:2022lak,Kuang:2023vac,Liu:2021rtn,Debastiani:2017msn,Jin:2020jfc,Zhang:2020hoh,Wang:2019rdo,liu:2020eha,Lu:2020cns,Giron:2020wpx,Karliner:2020dta,universe7040094,Dong:2025coi,Dosch:2020hqm,Chen:2024orv,NgaOngodo:2025mkt}. 
Many theoretical studies suggest that $X(6900)$ is a $2S$-wave fully charmed tetraquark state \cite{liu:2020eha,Lu:2020cns,Giron:2020wpx,Karliner:2020dta,universe7040094,Dong:2025coi}, while others propose it as an $1S$-wave state with total spin quantum number $S=2$ \cite{Dosch:2020hqm}.
For $X(6600)$, some works interpret it as a $2S$-wave state \cite{Dong:2025coi,NgaOngodo:2025mkt}, whereas others regard it as a $1S$-wave state \cite{liu:2020eha}.
There are also studies that suggest $X(7200)$ could be a $3S$-wave state \cite{Zhu:2020xni,universe7040094,Dong:2025coi} or a $2P$-wave state \cite{Yu:2022lak}.

In addition, some new experimental evidence for fully bottom tetraquark states has been reported.
In 2016, the CMS Collaboration announced the first observation of $\Upsilon(1S)$ pair production in $pp$ collisions at $\sqrt{s} = 8$ TeV \cite{CMS:2016liw}. 
They found an excess with a global significance of 3.6 $\sigma$ in the $4\mu$ channel. 
The invariant mass of this excess is $18.4\pm0.1\pm0.2$ GeV \cite{durgut2018aps}. 
Two years later, the LHCb Collaboration performed a research for the decay of the beautiful tetraquark, $X_{b\bar{b}b\bar{b}}$, to the $\Upsilon(1S)\mu^+\mu^-$ final state. 
However, no significant excess is seen for any mass hypothesis in the range $[17.5,20.0]$ GeV \cite{LHCb:2018uwm}.
In 2019, other production of double $\Upsilon(1S)$ is observed in Cu + Au collisions at the Relativistic Heavy Ion Collider by the A$_N$DY Collaboration. 
They showed evidence of a significant peak at $M=18.12 \pm 0.15 \pm 0.6$ GeV \cite{ANDY:2019bfn}, which is compatible with the first observation of an all-$b$ tetraquark. 
The fully bottom tetraquark states have also been the subject of extensive theoretical investigations \cite{Anwar:2017toa,Bai:2016int,Chen:2019dvd}.
Given that only a few experiments have observed these possible fully bottom tetraquark states, further experimental investigations are required to confirm their existence.

Fully heavy tetraquark states, especially the fully charmed and fully bottom tetraquark states, have attracted extensive attention from theorists. A wide range of theoretical approaches have been employed to study them, including the QCD sum rules \cite{Chen:2016jxd,Chen:2022sbf,Wang:2022xja}, a variety of quark models \cite{Faustov:2022mvs,Liu:2019zuc,Liu:2021rtn,Jin:2020jfc,Zhang:2022qtp,Wang:2021kfv,Meng:2024yhu}, the diquark models \cite{Dong:2022sef,Debastiani:2017msn,Giron:2020wpx,Mutuk:2021hmi}, the chromomagnetic model \cite{Weng:2020jao,Deng:2020iqw,Wu:2016vtq}, the diffusion Monte Carlo approach \cite{Bai:2016int,Gordillo:2020sgc,Ma:2022vqf}, and other phenomenological models \cite{Karliner:2020dta,universe7040094,Dong:2025coi,Dosch:2020hqm,NgaOngodo:2025mkt}. 
A few studies suggest that the fully heavy tetraquark states $cc\bar{c}\bar{c}$ and $bb\bar{b}\bar{b}$ could be bound states. In Ref. \cite{Wu:2016vtq}, the lowest mass of the $cc\bar{c}\bar{c}$ system lies below the di-$\eta_c$ threshold, while the lightest $S$-wave $bb\bar{b}\bar{b}$ state appears below the $\eta_b\eta_b$ threshold in Refs. \cite{Bedolla:2019zwg,Agaev:2023wua}. 
However, the majority of investigations conclude that the $S$-wave fully heavy tetraquarks are located above the corresponding double-meson thresholds. 
For example, in Ref. \cite{Hughes:2017xie}, the authors used the first-principles lattice nonrelativistic QCD methodology to study the low-lying spectrum of the $bb\bar{b}\bar{b}$ tetraquark states with quantum numbers $0^{++}$, $1^{+-}$ and $2^{++}$. 
Unfortunately, they found no evidence of a QCD bound fully heavy tetraquark below the lowest noninteracting thresholds in the channels studied.
It is evident that theorists hold diverse views regarding the new experimental observations, offering a variety of interpretations and even arriving at some contradictory conclusions. 
Meanwhile, some progress has been made in the study of mixed charm-bottom tetraquark states \cite{Mutuk:2022nkw,Agaev:2024pej,Wu:2024hrv}. 
Similarly, many issues remain unresolved.

Nowadays, a variety of new hadrons are continuously being observed in experiments, while theoretical progress remains relatively limited. 
Further theoretical studies are still required to deepen our understanding of the mechanisms underlying strong interactions. 
In this work, we adopted the diquark-antidiquark configuration to address the prohibition of light meson exchange in the fully heavy tetraquark state.
We then systematically investigated the mass spectra and decay channels of the fully heavy tetraquark states.

This paper is organized as follows. 
In Sec.~\ref{sec:theoretical model}, we present the effective Hamiltonian employed in this study and provide the flavor-color-spin basis wave functions corresponding to different quantum numbers. 
Subsequently, we introduce the formulas used for calculating the strong decay widths.
In Sec.~\ref{sec:results and discussion}, we present the model parameters required for the calculations and describe the methods for determining these parameters.
We then show the calculation results for the possible fully heavy tetraquark states and provide an analysis of their decay channels.
We summarize in Sec. \ref{sec:summary}.
The nature units $c = \hbar = 1$ are used throughout.

\section{Theoretical model \label{sec:theoretical model}}

\subsection{Effective Hamiltonian}

The bound state system formed by four fully heavy quarks can be seen as a unique example of direct interaction between quarks and gluons.
Consequently‌, the interaction in this system is mainly manifested as the one-gluon exchange process \cite{DeRujula:1975qlm,Guo:2021yws}.
For the fully heavy tetraquark systems, the improved nonrelativistic effective Hamiltonian reads \cite{Maiani:2004vq,Drenska:2008gr}

\begin{equation}
	H =\sum_{i=1}^4 m_i+H_{SS}+H_{SL}+H_L,\label{eq:Hamiltonian}
\end{equation}
where
\begin{equation}
	\begin{aligned}
		H_{SS}={}&\sum_{i<j}2\kappa_{ij}\left(\bm{S}_i\cdot\bm{S}_j\right),\\
		H_{SL}={}&2A\left(\bm{S}_{\mathcal{Q}}\cdot\bm{L}+\bm{S}_{\bar{\mathcal{Q}}^{\prime}}\cdot\bm{L}\right), \\
		H_L={}&B\frac{\bm{L}\left(\bm{L}+1\right)}{2}.
	\end{aligned}
\end{equation}

Here, $m_i$ and $\bm{S}_i$ denote the mass and spin of the $i$ th constituent quark, respectively.
The parameter $\kappa_{ij}$ is related to the strong coupling constant, the effective mass of the constituent quarks, the color coupling strength, and the spatial feature of the hadron.
$H_{SS}$ represents the spin-spin interaction of each quark pair in the tetraquark. 
$H_{SL}$ and $H_L$ describe spin-orbit coupling and pure orbital contribution, respectively. 
$\bm{S}_{\mathcal{Q}}$ in $H_{SL}$ represents the spin operator of the diquark configuration $\mathcal{Q} = [Q_1 Q_2]$. 
The orbital angular momentum $\bm{L}$ describes the relative motion between the diquark $\mathcal{Q}$ and the antidiquark $\bar{\mathcal{Q}}^{\prime}$. 
The coefficients $A$ and $B$ are phenomenological parameters to be determined from experimental data.

\subsection{Basis wave functions}

To obtain the mass spectra and decays of the fully heavy tetraquark states, we need to determine their wave functions. 
Here, we show only the basis wave functions of ground states, and we represent them as the symbol $|{(Q_1Q_2)_s^c(\bar{Q}_3\bar{Q}_4)_s^c}\rangle$, where the superscript denotes the color and the subscript denotes the spin.

Considering the Pauli principle, if the two quarks in a diquark have the same flavor, the spin and color of the diquark must have opposite symmetries. 
Therefore, we define an anti-Kronecker delta
\begin{equation}
	\bar{\delta}_{ij}=
	\left\{
	\begin{matrix}
		0\quad\text{if}\ i=j,\\
		1\quad\text{if}\ i\neq j.
	\end{matrix}\right.
\end{equation}

Now, the basis wave functions of the fully heavy tetraquark system based on flavor, spin and color can be constructed as follows.

\indent (i) For the quantum numbers $J^P=0^+$, the basis vectors $\alpha_i\ (i= 1, 2, 3, 4)$ can be labeled as
\begin{equation}
	\begin{aligned}
		\alpha_1=&|(Q_1Q_2)_0^{\bar{3}}(\bar{Q}_3\bar{Q}_4)_0^3\rangle\bar{\delta}_{12}\bar{\delta}_{34},\\
		\alpha_2=&|(Q_1Q_2)_0^6(\bar{Q}_3\bar{Q}_4)_0^{\bar{6}}\rangle,\\
		\alpha_3=&|(Q_1Q_2)_1^{\bar{3}}(\bar{Q}_3\bar{Q}_4)_1^3\rangle,\\
		\alpha_4=&|(Q_1Q_2)_1^6(\bar{Q}_3\bar{Q}_4)_1^{\bar{6}}\rangle\bar{\delta}_{12}\bar{\delta}_{34}.\\
	\end{aligned}
\end{equation}
\indent (ii) For the axial vector tetraquark states with quantum numbers $J^P=1^+$, the basis vectors $\beta_i\ (i= 1, 2,..., 6)$ can be expressed as
\begin{equation}
	\begin{aligned}
		\beta_1=&|(Q_1Q_2)_0^6(\bar{Q}_3\bar{Q}_4)_1^{\bar{6}}\rangle\bar{\delta}_{34},\\
		\beta_2=&|(Q_1Q_2)_0^{\bar{3}}(\bar{Q}_3\bar{Q}_4)_1^3\rangle\bar{\delta}_{12},\\
		\beta_3=&|(Q_1Q_2)_1^6(\bar{Q}_3\bar{Q}_4)_0^{\bar{6}}\rangle\bar{\delta}_{12},\\
		\beta_4=&|(Q_1Q_2)_1^{\bar{3}}(\bar{Q}_3\bar{Q}_4)_0^3\rangle\bar{\delta}_{34},\\
		\beta_5=&|(Q_1Q_2)_1^6(\bar{Q}_3\bar{Q}_4)_1^{\bar{6}}\rangle\bar{\delta}_{12}\bar{\delta}_{34},\\
		\beta_6=&|(Q_1Q_2)_1^{\bar{3}}(\bar{Q}_3\bar{Q}_4)_1^3\rangle.\\
	\end{aligned}
\end{equation}
\indent (iii) For $J^P=2^+$ states, the basis vectors $\gamma_i\ (i= 1, 2)$ of the fully heavy tetraquark states read
\begin{equation}
	\begin{aligned}
		\gamma_1=&|(Q_1Q_2)_1^6(\bar{Q}_3\bar{Q}_4)_1^{\bar{6}}\rangle\bar{\delta}_{12}\bar{\delta}_{34},\\
		\gamma_2=&|(Q_1Q_2)_1^{\bar{3}}(\bar{Q}_3\bar{Q}_4)_1^3\rangle.
	\end{aligned}
\end{equation}
The $\bar{\delta}_{12}$ (or $\bar{\delta}_{34}$) means that if $Q_1Q_2$ (or $\bar{Q}_3\bar{Q}_4$) have the same flavor, then this state cannot exist. 
We also consider the charge conjugation and excited states with $L=1$. The complete wave functions are listed in the Appendix.

\subsection{Decay width}

By inserting the basis vectors into the Hamiltonian (\ref{eq:Hamiltonian}), we can obtain the mass spectra and corresponding wave functions of the fully heavy tetraquark states.
We can then calculate the width of each tetraquark state decaying into the two meson final states.

For a particle of mass $M$ and four-momentum $P$ decaying into two particles with momenta $p_1$ and $p_2$, the partial decay rate can be written as
\begin{equation}
	d\Gamma=\frac{\left(2\pi\right)^4}{2M}|\mathcal{A}|^2d\Phi\left(P;p_1,p_2\right),
\end{equation}
where $\mathcal{A}$ is the amplitude connecting the initial state to the final state. And the symbol $d\Phi$ is an element of two-body phase space given by
\begin{equation}
	\begin{aligned}
		d\Phi\left(P;p_1,p_2\right)&=\delta^4\left(P-p_1-p_2\right)\\
		&\times\frac{d^3p_1}{\left(2\pi\right)^32E_1}\frac{d^3p_2}{\left(2\pi\right)^32E_2}.
	\end{aligned}
\end{equation}
In the center of momentum frame, considering four-momentum conservation in the decay process, we have
\begin{equation}
	|\bm{p}_1|=|\bm{p}_2|=\frac{1}{2M}\sqrt{\lambda\left(M^2,m_1^2,m_2^2\right)}
\end{equation}
where $\lambda\left(\alpha,\beta,\gamma\right)=\alpha^2+\beta^2+\gamma^2-2\alpha\beta-2\alpha\gamma-2\beta\gamma$ is the K\"{a}ll\'{e}n function. 
The notations $m_1$ and $m_2$ represent the masses of the subparticles in the final state of the two-body decay, respectively.
Thus, the partial decay rate can be written as
\begin{equation}
	d\Gamma=\frac{|\mathcal{A}|^2}{32\pi^2}\frac{|\bm{p}_1|}{M^2}d\Omega.
\end{equation}
Then, by integrating the differentiation above, we obtain
\begin{equation}
	\Gamma=\frac{|\mathcal{A}|^2|\bm{p}_1|}{8\pi M^2}.\label{eq:decay width}
\end{equation}

\section{results and discussion \label{sec:results and discussion}}

\subsection{Model parameters}

To calculate the ground-state masses and corresponding wave functions of fully heavy tetraquark states, we need to determine the constituent masses of the charm and bottom quarks, as well as the parameters $\kappa_{ij}$ associated with quark pairs.
Similar to the method in Refs. ~\cite{Kang:2025xqm,Liu:2025jyn}, we can obtain these model parameters by fitting experimental values of conventional hadrons.
For conventional hadrons that have not yet been observed experimentally, we used calculations based on lattice QCD \cite{Alexandrou:2017xwd,Brown:2014ena}.
The effective masses of the charm quark and the bottom quark can be obtained as $m_c=1556$ MeV and $m_b=4755$ MeV, respectively. 
Considering the differences in the effective masses of constituent quarks extracted from conventional ground-state hadrons, we introduce an uncertainty of  $\pm 0.5 \%$ for the effective masses of constituent quarks.

Without loss of generality, in the conventional hadron system, the color coupling between any two quarks can be only $1$ or $\bar{3}$, where $1$ corresponds to a pair of quark-antiquark forming a meson and $\bar{3}$ corresponds to the coupling of any two quarks in a baryon.
Thus, we first determine $(\kappa_{Q\bar{Q}})^1$ with mesons and $(\kappa_{QQ})^{\bar{3}}$ with baryons.

For mesons with $L=0$, Eq.~\hyperref[eq:Hamiltonian]{\eqref{eq:Hamiltonian}} provides the foundation for deriving 
\begin{equation}
	M=m_{Q_1}+m_{Q_2}+(\kappa_{Q_1\bar{Q}_2})^1\left[J\left(J+1\right)-\frac{3}{2}\right].
\end{equation}
With these equations, we can calculate the value of $(\kappa_{Q_1\bar{Q}_2})^1$. Their numerical results are tabulated in Table~\ref{tab:kappa}.
\begin{table}[htbp]
	\caption{ Parameter $\kappa_{ij}$ (in MeV) is used for $Q\bar{Q}$ with a singlet color configuration ($1$) and $QQ$ with an antitriplet color configuration ($\bar{3}$). Because of symmetry, these parameters satisfy $\kappa_{ij}=\kappa_{ji}$. And these parameters are derived from fitting the $1S$-wave mesons and baryons.}\label{tab:kappa}
	\begin{ruledtabular}
		\begin{tabular}{cccccc}
			$(\kappa_{c\bar{c}})^{1}$ & $(\kappa_{b\bar{b}})^{1}$ & $(\kappa_{b\bar{c}})^{1}$ & $(\kappa_{cc})^{\bar{3}}$ & $(\kappa_{bb})^{\bar{3}}$ & $(\kappa_{bc})^{\bar{3}}$\\
			\hline
			56 & 31 & 32 & 7 & 4 & 11\\
		\end{tabular}
	\end{ruledtabular}
\end{table}

To determine $(\kappa_{QQ})^{\bar{3}}$ with baryons, Eq.~\hyperref[eq:Hamiltonian]{\eqref{eq:Hamiltonian}} can be reformulated as
\begin{align}
	M=& \sum_{i=1}^{3}m_i+2\left(\kappa_{Q_1Q_2}\right)^{\bar{3}}\left(\bm{S}_{Q_1}\cdot\bm{S}_{Q_2}\right)+\left(\kappa_{Q_1Q_3}\right)^{\bar{3}}\left(\bm{S}_{Q_1}\cdot\bm{S}_{Q_3}\right)\notag\\
	+& \left(\kappa_{Q_2Q_3}\right)^{\bar{3}}\left(\bm{S}_{Q_2}\cdot\bm{S}_{Q_3}\right).\label{eq:kappa baryon}
\end{align}
For simplicity, we choose baryons that contain two quarks of the same flavor for our calculation.
In this case, Eq.~\hyperref[eq:kappa baryon]{\eqref{eq:kappa baryon}} can be written as
\begin{align}
	M=& 2m_{Q_1}+m_{Q_2}+\left(\kappa_{Q_1Q_1}\right)^{\bar{3}}\left[\bm{S}\left(\bm{S}+1\right)-\frac{3}{2}\right]\notag\\
	+& \left(\kappa_{Q_1Q_2}\right)^{\bar{3}}\left[J\left(J+1\right)-\bm{S}\left(\bm{S}+1\right)-\frac{3}{4}\right],
\end{align}
where $J$ is the total angular momentum quantum number of the baryon and $\bm{S}$ represents the total spin of the two identical-flavor quarks within the baryon.
Therefore, we can obtain the model parameters $\kappa^{\bar{3}}$ required for the calculation.
All the parameters are shown in Table~\ref{tab:kappa}. It is worth noting that the fitting results of these parameters $\kappa_{ij}$ have corresponding uncertainties. To obtain more reliable results, we consider the $\pm 10 \%$ difference of the parameters $\kappa_{ij}$ in the specific calculation.

For the tetraquark configuration $|(Q_1Q_2)(\bar{Q}_3\bar{Q}_4)\rangle$, its diquark-antidiquark color characteristic is a superposition of $|(Q_1Q_2)^{\bar{3}}(\bar{Q}_3\bar{Q}_4)^3\rangle$ and $|(Q_1Q_2)^{6}(\bar{Q}_3\bar{Q}_4)^{\bar{6}}\rangle$.
According to the hypothesis of strict Casimir scaling, in the configuration $|(Q_1Q_2)^{\bar{3}}(\bar{Q}_3\bar{Q}_4)^3\rangle$, the parameter $(\kappa_{Q\bar{Q}})^{8}=\frac{1}{4}(\kappa_{Q\bar{Q}})^1$.
Similarly, in the configuration $|(Q_1Q_2)^{6}(\bar{Q}_3\bar{Q}_4)^{\bar{6}}\rangle$, the parameters $(\kappa_{QQ})^6=-\frac{1}{2}(\kappa_{QQ})^{\bar{3}}$ and $(\kappa_{Q\bar{Q}})^{8}=\frac{5}{8}(\kappa_{Q\bar{Q}})^1$.

After studying the ground state of the fully heavy tetraquark state, we can include the contributions from $H_{SL}$ and $H_L$ to calculate the masses of the $L = 1$ excited states. 
By applying Eq.~\hyperref[eq:Hamiltonian]{\eqref{eq:Hamiltonian}} to charmonium mesons with $S = 1$ and $L = 0, 1$, specifically $J/\psi$, $\chi_{c0}$, $\chi_{c1}$, and $\chi_{c2}$, we can fit the coefficients in $H_{SL}$ and $H_L$. 
We get $A_{cc} = 21.83$ MeV and $B_{cc} = 426.06$ MeV. 
Similarly, using the bottomonium mesons $\Upsilon$, $\chi_{b0}$, $\chi_{b1}$, and $\chi_{b2}$, we can obtain $A_{bb} = 8.23$ MeV and $B_{bb} = 438.72$ MeV. 
For the mixed charm-bottom tetraquark states, we use the average values of $A_{cc}$ and $A_{bb}$ (as well as $B_{cc}$ and $B_{bb}$) for our calculations.

Since the tetraquark states discussed in this work are structured as diquark-antidiquark systems, quark rearrangement is required for them to strongly decay into two mesons. 
The decay widths are calculated using Eq.~\hyperref[eq:decay width]{\eqref{eq:decay width}} with $\mathcal{A}$ = 2600 MeV. 
The parameter $\mathcal{A}$ can be obtained by fitting the decay process of the scalar meson \cite{Maiani:2004uc}.
As higher-order decays (such as $P$-wave and $D$-wave decays) contribute less significantly to the total decay width, our analysis focuses primarily on $S$-wave decays. 
Of course, we also analyze and discuss $P$-wave decay modes in special cases.

\begin{figure*}[htbp]
	\centering
	\begin{minipage}{0.32\textwidth}
		\centering
		\includegraphics[width=1.0\textwidth,height=0.92\textwidth]{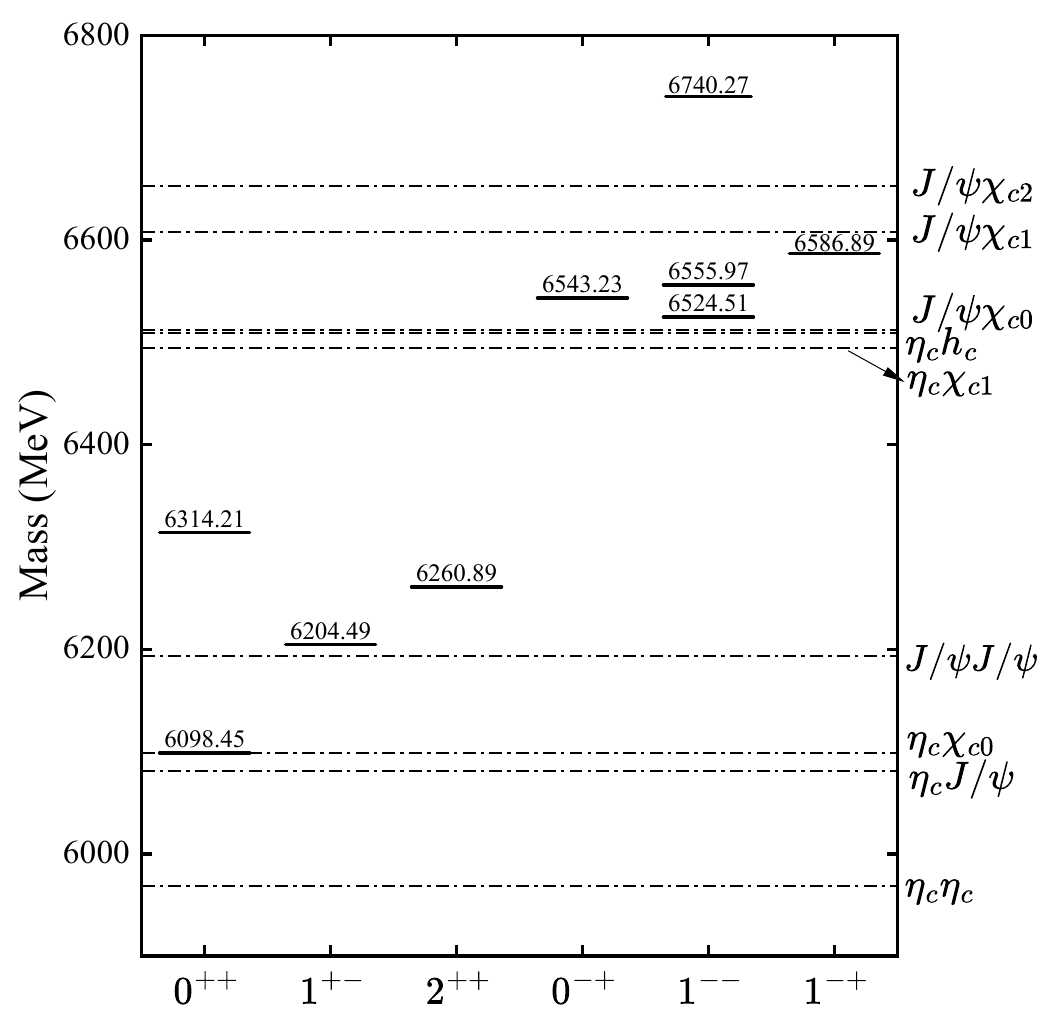}
		\text{(a) $[cc][\bar{c}\bar{c}]$}
	\end{minipage}\hfill
	\begin{minipage}{0.32\textwidth}
		\centering
		\includegraphics[width=1.0\textwidth,height=0.92\textwidth]{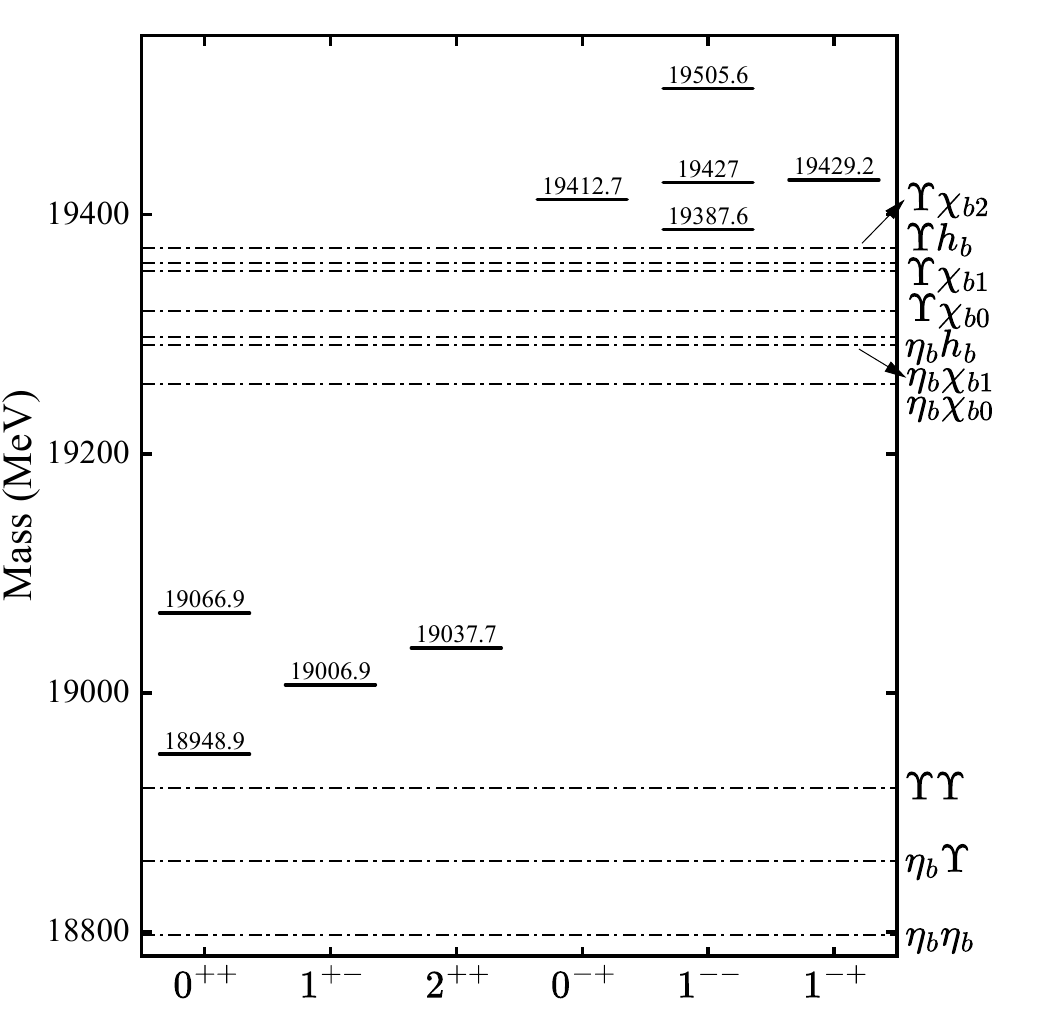}
		\text{(b) $[bb][\bar{b}\bar{b}]$}
	\end{minipage}\hfill
	\begin{minipage}{0.32\textwidth}
		\centering
		\includegraphics[width=1.0\textwidth,height=0.92\textwidth]{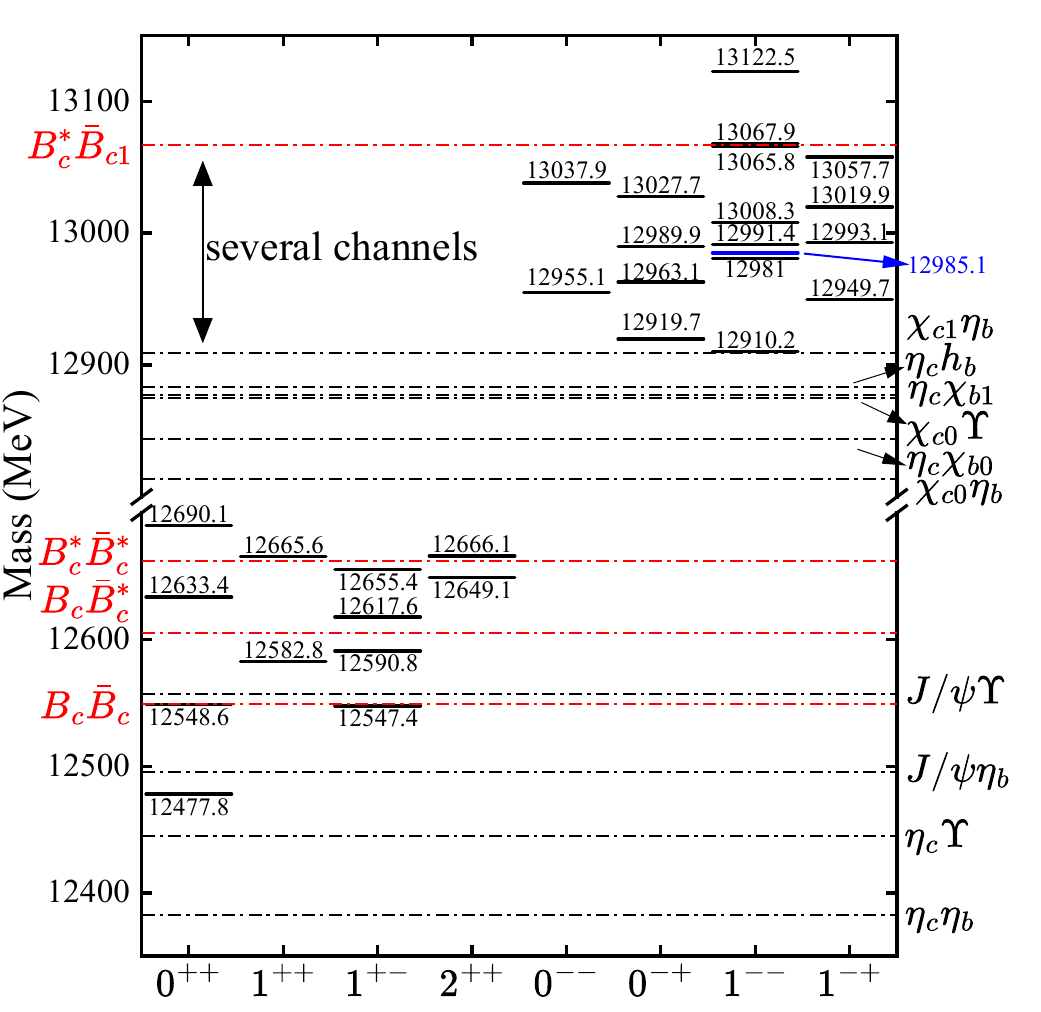}
		\text{(c) $[cb][\bar{c}\bar{b}]$}
	\end{minipage}
	\caption{The mass spectra of the fully heavy tetraquark states: (a) $[cc][\bar{c}\bar{c}]$, (b) $[bb][\bar{b}\bar{b}]$, and (c) $[cb][\bar{c}\bar{b}]$, each with different quantum numbers. The dash-dotted lines represent the corresponding meson-meson thresholds. In (c), the ``several channels'' include many threshold lines; for clarity and visual neatness, those lines overlapping with the energy levels have been omitted.}
	\label{fig:cccc bbbb cbcb}
\end{figure*}

\begin{figure*}[htbp]
	\centering
	\begin{minipage}{0.32\textwidth}
		\centering
		\includegraphics[width=1.0\textwidth,height=0.92\textwidth]{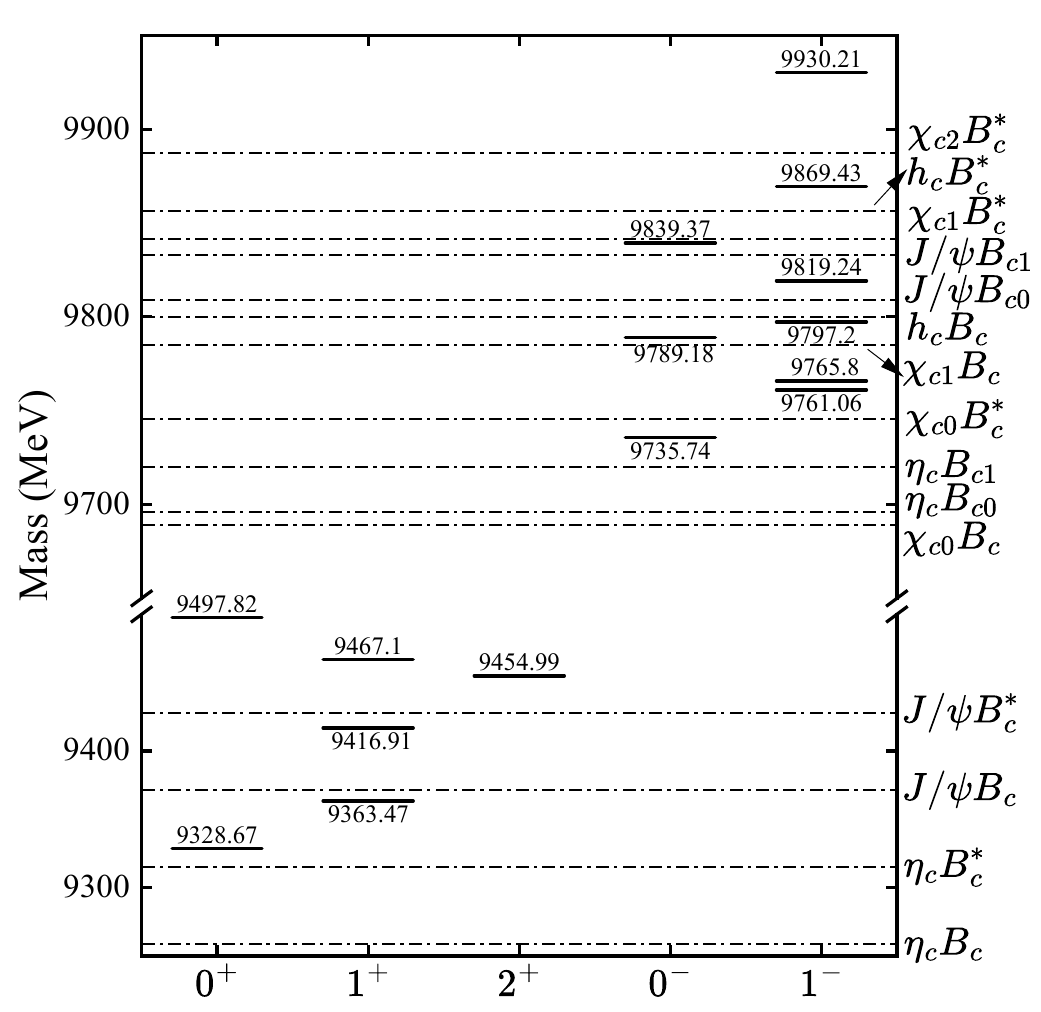}
		\text{(a) $[cc][\bar{c}\bar{b}]$}
	\end{minipage}\hfill
	\begin{minipage}{0.32\textwidth}
		\centering
		\includegraphics[width=1.0\textwidth,height=0.92\textwidth]{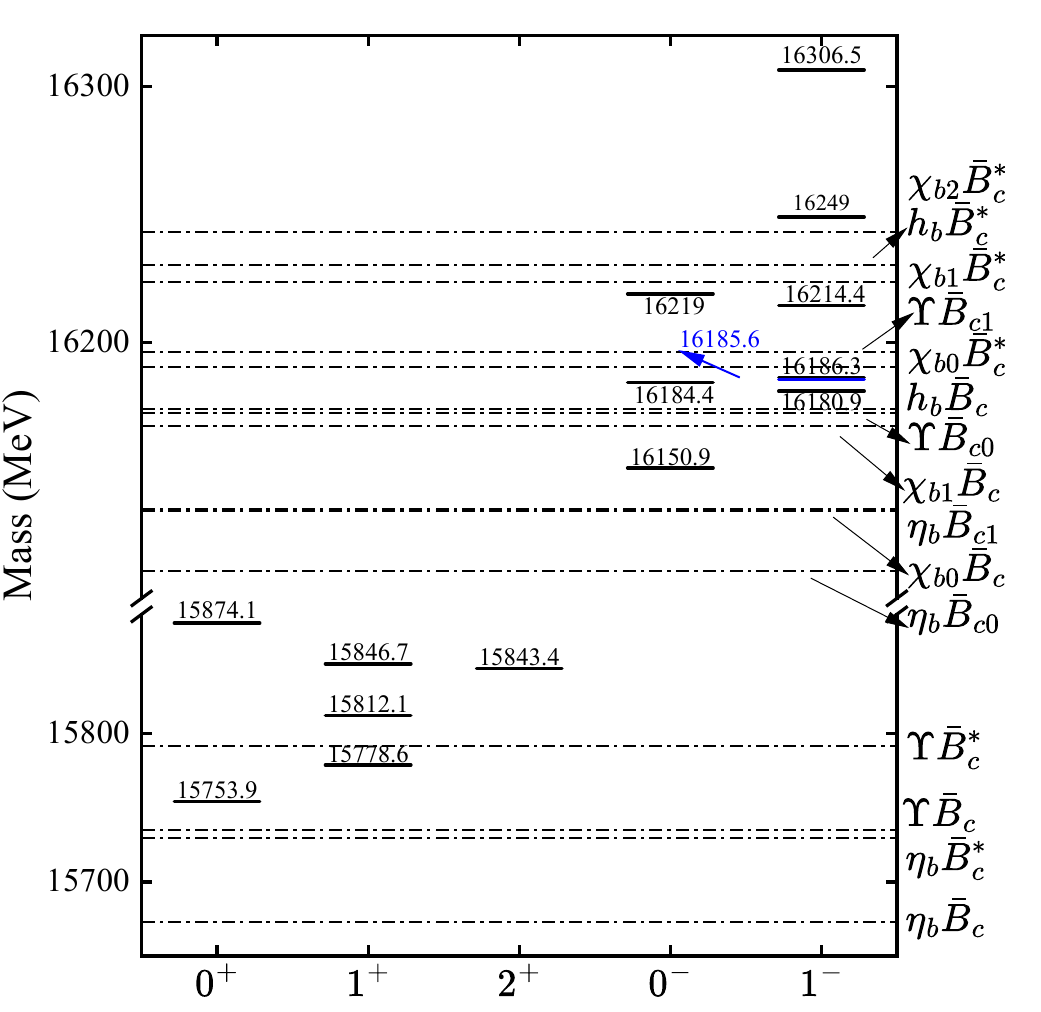}
		\text{(b) $[bb][\bar{b}\bar{c}]$}
	\end{minipage}\hfill
	\begin{minipage}{0.32\textwidth}
		\centering
		\includegraphics[width=1.0\textwidth,height=0.92\textwidth]{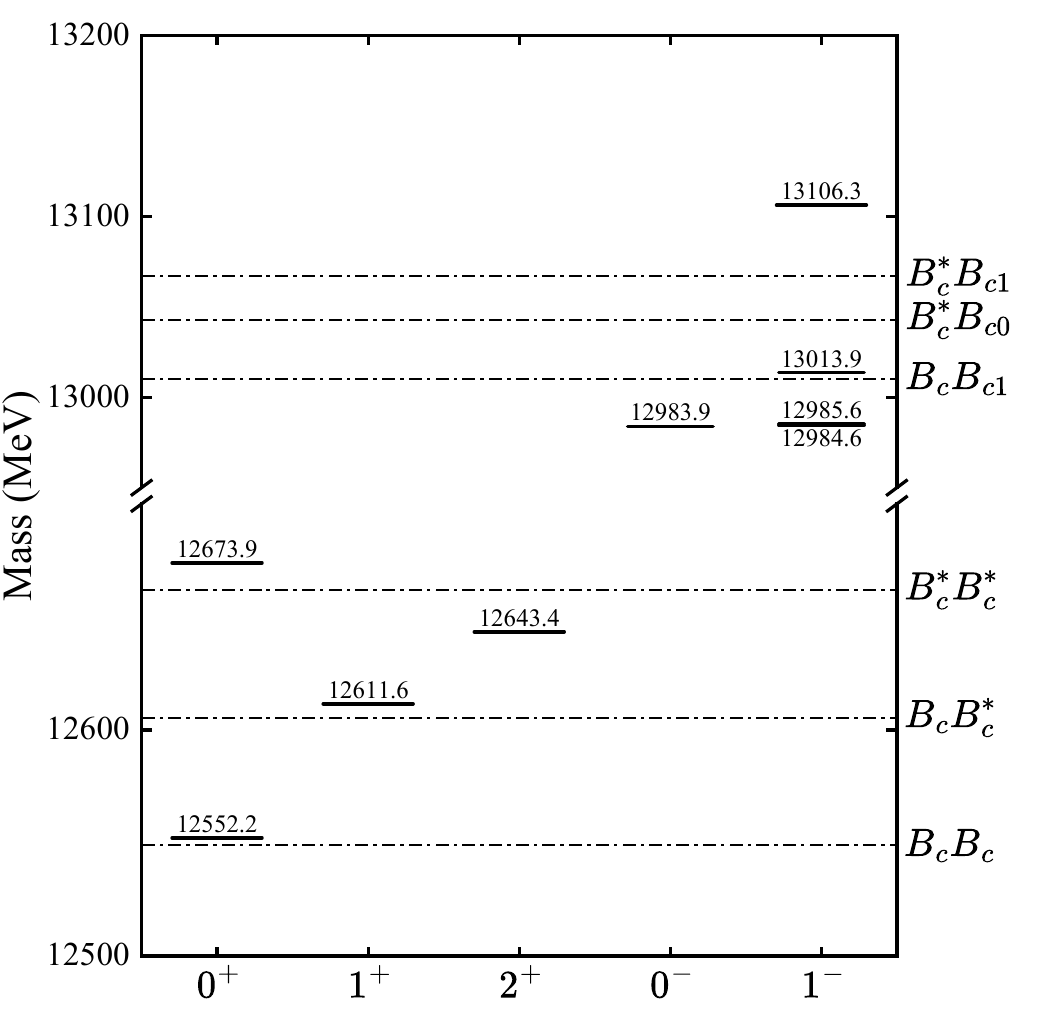}
		\text{(c) $[cc][\bar{b}\bar{b}]$}
	\end{minipage}
	\caption{The mass spectra of the fully heavy tetraquark states: (a) $[cc][\bar{c}\bar{b}]$, (b) $[bb][\bar{b}\bar{c}]$, and (c) $[cc][\bar{b}\bar{b}]$, each with different quantum numbers. The dash-dotted lines represent the corresponding meson-meson thresholds.}
	\label{fig:cccb bbbc ccbb}
\end{figure*}

\begin{table*}[htbp]
	\caption{The quantum numbers, masses, eigenvectors of the basis wave functions, and decay channels of the fully heavy tetraquark $[cc][\bar{c}\bar{c}]$ and $[bb][\bar{b}\bar{b}]$ systems.}\label{tab:cccc bbbb masses and decay}
	\begin{ruledtabular}
		\begin{tabular}{ccccl}
			System & $J^{PC}$ & Mass (MeV) & Eigenvector & Decay channels [$\Gamma$ (MeV)]\\
			\hline
			$[cc][\bar{c}\bar{c}]$ & $0^{++}$ & 6098.45$\pm$18 & $(0.60, 0.80)$  & $\eta_c\eta_c[4.53]$\\\vspace{3mm}
			&& 6314.21$\pm$41 & $(-0.80, 0.60)$  & $\eta_c\eta_c[6.95];J/\psi J/\psi[4.14]$\\\vspace{3mm}
			& $1^{+-}$ & 6204.49$\pm$29 & 1 & $\eta_cJ/\psi[4.31]$\\\vspace{3mm}
			& $2^{++}$ & 6260.89$\pm$36 & 1 & $J/\psi J/\psi[3.14]$\\\vspace{3mm}
			& $0^{-+}$ & 6543.23$\pm$29 & 1 & $\eta_c\chi_{c0}[4.29]$\\
			& $1^{--}$ & 6524.51$\pm$18 & $(0.60, 0.80)$ & $\eta_ch_c[1.39];J/\psi\chi_{c0}[1.29]$\\
			&& 6740.27$\pm$41 & $(-0.80,0.60)$ & $\eta_ch_c[5.16];J/\psi\chi_{c0}[5.15];J/\psi\chi_{c1}[3.93];J/\psi\chi_{c2}[3.19]$\\\vspace{3mm}
			&& 6555.97$\pm$36 & 1 & $\eta_ch_c[2.43];J/\psi\chi_{c0}[2.38]$\\\vspace{3mm}
			& $1^{-+}$ & 6586.89$\pm$29 & 1 & $\eta_c\chi_{c1}[3.39]$\\
			$[bb][\bar{b}\bar{b}]$ & $0^{++}$ & 18948.9$\pm$110 & $(0.601, 0.799)$ & $\eta_b\eta_b[0.90];\Upsilon\Upsilon[0.39]$\\\vspace{3mm}
			&& 19066.9$\pm$121 & $(-0.799,0.601)$ & $\eta_b\eta_b[1.18];\Upsilon\Upsilon[0.87]$\\\vspace{3mm}
			& $1^{+-}$ & 19006.9$\pm$115 & 1 & $\eta_b\Upsilon[0.88]$\\\vspace{3mm}
			& $2^{++}$ & 19037.7$\pm$118 & 1 & $\Upsilon\Upsilon[0.78]$\\\vspace{3mm}
			& $0^{-+}$ & 19412.7$\pm$115 & 1 & $\eta_b\chi_{b0}[0.87];\Upsilon h_b[0.51]$\\
			& $1^{--}$ & 19387.6$\pm$110 & $(0.601, 0.799)$ & $\eta_bh_b[0.67];\Upsilon\chi_{b0}[0.58];\Upsilon\chi_{b1}[0.41];\Upsilon\chi_{b2}[0.27]$\\
			& & 19505.6$\pm$121 & $(-0.799,0.601)$ & $\eta_bh_b[1.00];\Upsilon\chi_{b0}[0.95];\Upsilon\chi_{b1}[0.86];\Upsilon\chi_{b2}[0.80]$\\\vspace{3mm}
			& & 19427.0$\pm$118 & 1 & $\eta_bh_b[0.80];\Upsilon\chi_{b0}[0.73];\Upsilon\chi_{b1}[0.60];\Upsilon\chi_{b2}[0.52]$\\
			& $1^{-+}$ & 19429.2$\pm$115 & 1 & $\eta_b$$\chi_{b1}[0.82];\Upsilon h_b[0.58]$\\
		\end{tabular}
	\end{ruledtabular}
\end{table*}

\begin{table*}[htbp]
	\caption{The quantum numbers, masses, eigenvectors of the basis wave functions, and decay channels of the fully heavy tetraquark $[cb][\bar{c}\bar{b}]$ system. 
		     For some tetraquark states, there is no available decay channel that both satisfies the conservation law and has a mass higher than the corresponding meson-meson threshold.}\label{tab:cbcb masses and decay}
	\begin{ruledtabular}
		\begin{tabular}{cccl}
			$J^{PC}$ & Mass (MeV) & Eigenvector & Decay channels [$\Gamma$ (MeV)]\\
			\hline
			$0^{++}$ & 12477.8$\pm$59 & $(-0.15,-0.52,-0.82,-0.20)$  & $\eta_c\eta_b[1.14]$\\
			& 12548.6$\pm$67 & $(0.57,-0.20,-0.17,0.78)$ & $\eta_c\eta_b[1.49]$\\
			& 12690.1$\pm$80 & $(-0.81,-0.06,0.04,0.59)$ & $\eta_c\eta_b[2.00];J/\psi\Upsilon[1.32];B_c\bar{B}_c[1.58];B_c^*\bar{B}_c^*[0.70]$\\\vspace{3mm}
			& 12633.4$\pm$75 & $(-0.02,0.83,-0.55,0.10)$ & $\eta_c\eta_b[1.82];J/\psi\Upsilon[1.01];B_c\bar{B}_c[1.23]$\\
			$1^{++}$ & 12665.6$\pm$78 & $(-0.79,-0.61)$ & $J/\psi\Upsilon[1.20];B_c\bar{B}_c^*[1.03];B_c^*\bar{B}_c^*[0.25]$\\\vspace{3mm}
			& 12582.8$\pm$70 & $(0.61,-0.79)$ & $J/\psi\Upsilon[0.59]$\\
			$1^{+-}$ & 12547.4$\pm$66 & $(0.46,0.48,-0.71,-0.21)$ & $\eta_c\Upsilon[1.17];J/\psi\eta_b[0.84]$\\
			& 12655.4$\pm$77 & $(0.77,-0.62,0.04,0.15)$ & $\eta_c\Upsilon[1.66];J/\psi\eta_b[1.46];B_c\bar{B}_c^*[0.94]$\\
			& 12590.8$\pm$71 & $(0.43,0.53,0.70,-0.20)$ & $\eta_c\Upsilon[1.39];J/\psi\eta_b[1.13]$\\\vspace{3mm}
			& 12617.6$\pm$73 & $(0.07,0.32,-0.01,0.94)$ & $\eta_c\Upsilon[1.51];J/\psi\eta_b[1.28];B_c\bar{B}_c^*[0.47]$\\
			$2^{++}$ & 12666.1$\pm$79 & $(-0.91,-0.41)$ & $J/\psi\Upsilon[1.20];B_c^*\bar{B}_c^*[0.27]$\\\vspace{3mm}
			& 12649.1$\pm$76 & $(0.41,-0.91)$ & $J/\psi\Upsilon[1.11]$\\
			$0^{--}$ & 13037.9$\pm$78 & $(-0.79,-0.61)$ & $J/\psi\chi_{b1}[0.76];\chi_{c1}\Upsilon[0.93];B_c\bar{B}_{c0}[0.91]$\\\vspace{3mm}
			& 12955.1$\pm$70 & $(0.61,-0.79)$ & \\
			$0^{-+}$ & 12919.7$\pm$66 & $(0.46,0.48,-0.71,-0.21)$ & $\eta_c\chi_{b0}[0.96];\chi_{c0}\eta_b[1.18]$\\
			& 13027.7$\pm$77 & $(0.77,-0.62,0.04,0.15)$ & $\eta_c\chi_{b0}[1.47];\chi_{c0}\eta_b[1.66];J/\psi h_b[0.61];h_c\Upsilon[0.74];B_c\bar{B}_{c0}[0.82]$\\
			& 12963.1$\pm$71 & $(0.43,0.53,0.70,-0.20)$ & $\eta_c\chi_{b0}[1.19];\chi_{c0}\eta_b[1.39]$ \\\vspace{3mm}
			& 12989.9$\pm$73 & $(0.07,0.32,-0.01,0.94)$ & $\eta_c\chi_{b0}[1.31];\chi_{c0}\eta_b[1.51]$ \\
			$1^{--}$ & 12910.2$\pm$59 & $(-0.15,-0.52,-0.82,-0.20)$ & $\eta_c\eta_b[0.57];\chi_{c0}\Upsilon[0.68]$\\
			& 12981$\pm$67 & $(0.57,-0.20,-0.17,0.78)$ & $\eta_c\eta_b[1.66];J/\psi\chi_{b0}[1.39];\chi_{c0}\Upsilon[1.76];\chi_{c1}\Upsilon[1.38];h_c\eta_b[1.59]$\\
			& 13122.5$\pm$80 & $(-0.81,-0.06,0.04,0.59)$ & $\eta_c\eta_b[1.66];J/\psi\chi_{b0}[1.39];J/\psi\chi_{b1}[1.21];J/\psi\chi_{b2}[1.15];\chi_{c0}\Upsilon[1.76];\chi_{c1}\Upsilon[1.38]$\\
			&&& $\chi_{c2}\Upsilon[1.16];h_c\eta_b[1.59];B_c\bar{B}_{c1}[1.34];B_c^*\bar{B}_{c0}[1.13];B_c^*\bar{B}_{c1}[0.94]$\\
			& 13065.8$\pm$75 & $(-0.02,0.83,-0.55,0.10)$ & $\eta_c\eta_b[1.45];J/\psi\chi_{b0}[1.14];J/\psi\chi_{b1}[0.91];J/\psi\chi_{b2}[0.82];\chi_{c0}\Upsilon[1.55];\chi_{c1}\Upsilon[1.10]$\\
			&&& $\chi_{c2}\Upsilon[0.80];h_c\eta_b[1.35];B_c\bar{B}_{c1}[0.95];B_c^*\bar{B}_{c0}[0.61]$\\
			& 13067.9$\pm$78 & $(-0.79,-0.61)$ & $\eta_c\eta_b[1.46];J/\psi\chi_{b0}[1.15];J/\psi\chi_{b1}[0.92];J/\psi\chi_{b2}[0.83];\chi_{c0}\Upsilon[1.56];\chi_{c1}\Upsilon[1.11]$\\
			&&& $\chi_{c2}\Upsilon[0.81];h_c\eta_b[1.36];B_c\bar{B}_{c1}[0.96];B_c^*\bar{B}_{c0}[0.63];B_c^*\bar{B}_{c1}[0.12]$\\
			& 12985.1$\pm$70 & $(0.61,-0.79)$ & $\eta_c\eta_b[1.09];J/\psi\chi_{b0}[0.59];\chi_{c0}\Upsilon[1.19];\chi_{c1}\Upsilon[0.43];h_c\eta_b[0.89]$\\
			& 13008.3$\pm$79 & $(-0.91,-0.41)$ & $\eta_c\eta_b[1.21];J/\psi\chi_{b0}[0.79];J/\psi\chi_{b1}[0.38];\chi_{c0}\Upsilon[1.31];\chi_{c1}\Upsilon[0.70];h_c\eta_b[1.05]$\\\vspace{3mm}
			& 12991.4$\pm$76 & $(0.41,-0.91)$ & $\eta_c\eta_b[1.13];J/\psi\chi_{b0}[0.65];\chi_{c0}\Upsilon[1.22];\chi_{c1}\Upsilon[0.51];h_c\eta_b[0.94]$\\
			$1^{-+}$ & 12949.7$\pm$66 & $(0.46,0.48,-0.71,-0.21)$ & $\eta_c\chi_{b1}[0.93];\chi_{c1}\eta_b[0.73]$\\
			& 13057.7$\pm$77 & $(0.77,-0.62,0.04,0.15)$ & $\eta_c\chi_{b1}[1.45];J/\psi h_b[0.85];\chi_{c1}\eta_b[1.38];h_c\Upsilon[0.96];B_c\bar{B}_{c1}[0.87];B_c^*\bar{B}_{c0}[0.49]$\\
			& 12993.1$\pm$71 & $(0.43,0.53,0.70,-0.20)$ & $\eta_c\chi_{b1}[1.17];\chi_{c1}\eta_b[1.05];h_c\Upsilon[0.31]$\\
			& 13019.9$\pm$73 & $(0.07,0.32,-0.01,0.94)$ & $\eta_c\chi_{b1}[1.29];J/\psi h_b[0.53];\chi_{c1}\eta_b[1.20];h_c\Upsilon[0.67];B_c\bar{B}_{c1}[0.39]$\\
		\end{tabular}
	\end{ruledtabular}
\end{table*}

\begin{table*}[htbp]
	\caption{The quantum numbers, masses, eigenvectors of the basis wave functions, and decay channels of the $[cc][\bar{c}\bar{b}]$ system.}\label{tab:cccb masses and decay}
	\begin{ruledtabular}
		\begin{tabular}{cccl}
			$J^{P}$ & Mass (MeV) & Eigenvector & Decay channels [$\Gamma$ (MeV)] \\
			\hline
			$0^{+}$ & 9328.67$\pm$43 & $(0.597, 0.803)$ & $\eta_cB_c[1.65]$\\\vspace{3mm}
			& 9497.82$\pm$60 & $(-0.803,0.597)$ & $\eta_cB_c[2.96];J/\psi B_c^*[1.61]$ \\
			$1^{+}$ & 9363.47$\pm$47 & $(0.622,-0.690,0.369)$ & $J/\psi B_c[1.36]$\\
			& 9467.10$\pm$58 & $(0.774,0.613,-0.160)$ & $\eta_cB_c^*[1.90];J/\psi B_c[2.37];J/\psi B_c^*[1.21]$\\\vspace{3mm}
			& 9416.91$\pm$52 & $(-0.116,0.385,0.916)$  & $\eta_cB_c^*[1.32];J/\psi B_c[1.96]$\\\vspace{3mm}
			$2^{+}$ & 9454.99$\pm$56 & 1 & $J/\psi B_c^*[1.01]$\\
			$0^{-}$ & 9735.74$\pm$47 & $(0.622,-0.690,0.369)$ & $\eta_cB_{c0}[0.29];J/\psi B_{c1}[0.15]$\\
			& 9839.37$\pm$58 & $(0.774,0.613,-0.160)$ & $\eta_cB_{c0}[2.28];J/\psi B_{c1}[2.15];\chi_{c0}B_c[0.46]$\\\vspace{3mm}
			& 9789.18$\pm$52 & $(-0.116,0.385,0.916)$ & $\eta_cB_{c0}[1.87];J/\psi B_{c1}[1.75]$\\
			$1^{-}$ & 9761.06$\pm$43 & $(0.597, 0.803)$ & $J/\psi B_{c1}[0.74];h_cB_c^*[1.16]$\\
			& 9930.21$\pm$60 & $(-0.803,0.597)$ & $\eta_cB_{c1}[2.22];J/\psi B_{c0}[2.10];J/\psi B_{c1}[2.48];\chi_{c0}B_c^*[1.73];\chi_{c1}B_c[1.58]$\\
			&&&$\chi_{c1}B_c^*[1.21];h_cB_c[1.97];h_cB_c^*[2.57];\chi_{c2}B_c^*[1.76]$\\
			& 9765.8$\pm$47 & $(0.622,-0.690,0.369)$ & $J/\psi B_{c1}[0.84];h_cB_c^*[1.23]$\\
			& 9869.43$\pm$58 & $(0.774,0.613,-0.160)$ & $\eta_cB_{c1}[1.71];J/\psi B_{c0}[1.55];J/\psi B_{c1}[2.05];\chi_{c0}B_c^*[0.98];\chi_{c1}B_c[0.67]$\\
			&&&$h_cB_c[1.40];h_cB_c^*[2.18];\chi_{c2}B_c^*[1.09]$\\
			& 9819.24$\pm$52 & $(-0.116,0.385,0.916)$ & $\eta_cB_{c1}[1.09];J/\psi B_{c0}[0.83];J/\psi B_{c1}[1.60];h_cB_c[0.58];h_cB_c^*[1.79]$\\
			& 9797.2$\pm$56 & 1 & $\eta_cB_{c1}[0.65];J/\psi B_{c1}[1.34];h_cB_c^*[1.59]$\\
		\end{tabular}
	\end{ruledtabular}
\end{table*}

\begin{table*}[htbp]
	\caption{The quantum numbers, masses, eigenvectors of the basis wave functions, and decay channels of the $[bb][\bar{b}\bar{c}]$ system.}\label{tab:bbbc masses and decay}
	\begin{ruledtabular}
		\begin{tabular}{cccl}
			$J^P$ & Mass (MeV) & Eigenvector & Decay channels [$\Gamma$ (MeV)] \\
			\hline
			$0^+$ & 15753.9$\pm$89 & $(0.595,0.803)$ & $\eta_b\bar{B}_c[0.85]$\\\vspace{3mm}
			& 15874.1$\pm$100 & $(-0.803,0.595)$ & $\eta_b\bar{B}_c[1.32];\Upsilon\bar{B}_c*[0.85]$\\
			$1^+$ & 15778.6$\pm$91 & $(0.624,-0.782,-0.02)$ & $\eta_b\bar{B}_c^*[0.66];\Upsilon\bar{B}_c[0.62]$\\
			& 15846.7$\pm$98 & $(0.782,0.624,0.010)$ & $\eta_b\bar{B}_c^*[1.01];\Upsilon\bar{B}_c[0.99];\Upsilon\bar{B}_c^*[0.69]$\\\vspace{3mm}
			& 15812.1$\pm$94 & $(0.005,-0.022,0.999)$ & $\eta_b\bar{B}_c^*[0.85];\Upsilon\bar{B}_c[0.82];\Upsilon\bar{B}_c^*[0.43]$\\\vspace{3mm}
			$2^+$ & 15843.4$\pm$97 & 1 & $\Upsilon\bar{B}_c^*[0.67]$\\
			$0^-$ & 16150.9$\pm$91 & $(0.624,-0.782,-0.02)$ & $\eta_b\bar{B}_{c0}[0.58];\chi_{b0}\bar{B}_c[0.37]$\\
			& 16219$\pm$98 & $(0.782,0.624,0.010)$ & $\eta_b\bar{B}_{c0}[0.94];\Upsilon\bar{B}_{c1}[0.43];\chi_{b0}\bar{B}_c[0.83]$\\\vspace{3mm}
			& 16184.4$\pm$94 & $(0.005,-0.022,0.999)$ & $\eta_b\bar{B}_{c0}[0.78];\chi_{b0}\bar{B}_c[0.64]$\\
			$1^-$& 16186.3$\pm$89 & $(0.595,0.803)$ & $\eta_b\bar{B}_{c1}[0.65];\Upsilon\bar{B}_{c0}[0.34];\chi_{b1}\bar{B}_c[0.39];h_b\bar{B}_c[0.32]$\\
			& 16306.5$\pm$100 & $(-0.803,0.595)$ & $\eta_b\bar{B}_{c1}[1.18];\Upsilon\bar{B}_{c0}[1.04];\Upsilon\bar{B}_{c1}[0.94];\chi_{b0}\bar{B}_c^*[0.96];\chi_{b1}\bar{B}_c[1.05]$\\
			&&&$\chi_{b1}\bar{B}_c^*[0.81];h_b\bar{B}_c[1.02];h_b\bar{B}_c^*[0.78];\chi_{b2}\bar{B}_c^*[0.71]$\\
			& 16180.9$\pm$91 & $(0.624,-0.782,-0.02)$ & $\eta_b\bar{B}_{c1}[0.62];\Upsilon\bar{B}_{c0}[0.27];\chi_{b1}\bar{B}_c[0.33];h_b\bar{B}_c[0.24]$\\
			& 16249$\pm$98 & $(0.782,0.624,0.010)$ & $\eta_b\bar{B}_{c1}[0.97];\Upsilon\bar{B}_{c0}[0.79];\Upsilon\bar{B}_{c1}[0.66];\chi_{b0}\bar{B}_c^*[0.69];\chi_{b1}\bar{B}_c[0.81]$\\
			&&&$\chi_{b1}\bar{B}_c^*[0.45];h_b\bar{B}_c[0.78];h_b\bar{B}_c^*[0.39];\chi_{b2}\bar{B}_c^*[0.22]$\\
			& 16214.4$\pm$94 & $(0.005,-0.022,0.999)$ & $\eta_b\bar{B}_{c1}[0.81];\Upsilon\bar{B}_{c0}[0.59];\Upsilon\bar{B}_{c1}[0.39];\chi_{b0}\bar{B}_c^*[0.44];\chi_{b1}\bar{B}_c[0.62];h_b\bar{B}_c[0.57]$\\
			& 16185.6$\pm$97 & 1 & $\eta_b\bar{B}_{c1}[0.65];\Upsilon\bar{B}_{c0}[0.33];\chi_{b1}\bar{B}_c[0.39];h_b\bar{B}_c[0.31]$
		\end{tabular}
	\end{ruledtabular}
\end{table*}

\begin{table*}[htbp]
	\caption{The quantum numbers, masses, eigenvectors of the basis wave functions, and decay channels of the fully heavy tetraquark $[cc][\bar{b}\bar{b}]$ system. 
		Similarly, for certain states, there is no available decay channel that both satisfies the conservation law and has a mass higher than the corresponding meson-meson threshold.}\label{tab:ccbb masses and decay}
	\begin{ruledtabular}
		\begin{tabular}{cccl}
			$J^P$ & Mass (MeV) & Eigenvector & Decay channels [$\Gamma$ (MeV)] \\
			\hline
			$0^+$ & 12552.2$\pm$67 & $(0.598, 0.801)$ & $B_c\bar{B}_c[0.24]$\\\vspace{3mm}
			& 12673.9$\pm$79 & $(-0.801,0.598)$ & $B_c\bar{B}_c[1.49];B_c^*\bar{B}_c^*[0.46]$\\\vspace{3mm}
			$1^+$ & 12611.6$\pm$73 & 1 & $B_c\bar{B}_c^*[0.33]$\\\vspace{3mm}
			$2^+$ & 12643.4$\pm$76 & 1 & \\\vspace{3mm}
			$0^-$ & 12983.9$\pm$73 & 1 & \\
			$1^-$ & 12984.6$\pm$67 & $(0.598, 0.801)$ & \\
			& 13106.3$\pm$79 & $(-0.801,0.598)$ & $B_c\bar{B}_{c1}[1.24];B_c^*\bar{B}_{c0}[1.01];B_c^*\bar{B}_{c1}[0.79]$\\
			& 13013.9$\pm$73 & 1 & $B_c\bar{B}_{c1}[0.24]$\\
			& 12985.6$\pm$76 & 1 & \\
		\end{tabular}
	\end{ruledtabular}
\end{table*}

\subsection{$[cc][\bar{c}\bar{c}]$, $[bb][\bar{b}\bar{b}]$, and $[cb][\bar{c}\bar{b}]$ systems}

From the basis wave functions of tetraquark $[cc][\bar{c}\bar{c}]$, $[bb][\bar{b}\bar{b}]$, and $[cb][\bar{c}\bar{b}]$ configurations, it can be seen that these states possesses a well-defined charge conjugation parity ($C$ parity). 
For these states, conservation of $C$ parity must be taken into account during the decay process. 
Their mass spectra and decay channels were shown in Fig.~\ref{fig:cccc bbbb cbcb} and Tables ~\ref{tab:cccc bbbb masses and decay} and \ref{tab:cbcb masses and decay}. 
The two final-state mesons we choose for these decays are in the $1S$ and $1P$ states. 
To date, only the spin-0 states of the $B_c(1S)$ and $B_c(2S)$ mesons have been observed experimentally. 
Therefore, for the remaining $c\bar{b}$ mesons in the $1S$ and $1P$ states, we use the masses predicted by lattice QCD calculations \cite{Mathur:2018epb}. 
For simplicity, we denote the $c\bar{b}$ mesons with quantum numbers $J^{P} = 0^+$ and $1^+$ as $B_{c0}$ and $B_{c1}$, respectively.

\subsubsection{$[cc][\bar{c}\bar{c}]$ system}

For the tetraquark state $[cc][\bar{c}\bar{c}]$ with different quantum numbers, the mass spectra and corresponding meson-meson thresholds are exhibited in Fig.~\ref{fig:cccc bbbb cbcb}(a) and the decay channels are shown in Table~\ref{tab:cccc bbbb masses and decay}.

The $1S$-wave fully charmed tetraquark states possess quantum numbers $J^{PC} = 0^{++}$, $1^{+-}$, and $2^{++}$.  
There are two states with quantum numbers $0^{++}$, and their masses are 6098.45, and 6314.21 MeV, respectively.
The lower-mass state, at 6098.45 MeV, lies above both the $\eta_c\eta_c$ and $\eta_c J/\psi$ thresholds. 
However, due to conservation laws, it can decay naturally only into the $\eta_c\eta_c$ channel, while decay into $\eta_c J/\psi$ is forbidden. 
The higher-mass state, at 6314.21 MeV, lies above both $\eta_c\eta_c$ and $J/\psi J/\psi$ thresholds and is allowed to decay into both of these two final states.
However, the narrower width of the $J/\psi J/\psi$ final state suggests that this state is more likely to be found in the $J/\psi J/\psi$ invariant mass spectrum. 
Furthermore, the wide width of the decay to $\eta_c\eta_c$ indicates that this state is difficult to distinguish from the background.

For the fully charmed tetraquark states with quantum numbers $1^{+-}$ and $2^{++}$, the mass spectra are 6204.49 and 6260.89 MeV, respectively. 
Both lie above the $\eta_c J/\psi$ and $J/\psi J/\psi$ thresholds. 
Nevertheless, due to quantum number constraints, the $1^{+-}$ state can decay only into the $\eta_c J/\psi$ channel. 
The state with quantum numbers $2^{++}$ is allowed to decay solely into the final-state mesons $J/\psi J/\psi$.
In addition, the narrow width of the decay to $J/\psi J/\psi$ indicates that the state can be found within the invariant mass spectrum of $J/\psi J/\psi$.
This suggests that the fully charmed tetraquark state with quantum number $2^{++}$ at 6260.89 MeV may be $X(6200)$ \cite{Song:2024ykq}, which has been confirmed in recent experimental analyses.
Unfortunately, we did not find reliable candidates for $X(6600)$, $X(6900)$, and $X(7200)$ in the $1S$-wave fully charmed tetraquark system.

The $P$-wave fully charmed tetraquark states have quantum numbers $J^{PC} = 0^{-+}$, $1^{--}$, and $1^{-+}$. 
For the quantum numbers $0^{-+}$, there exists only one tetraquark state with a mass of 6543.23 MeV. 
Clearly, the feature of this state being above the $\eta_c\chi_{c0}$ threshold allows it to naturally decay into two mesons $\eta_c\chi_{c0}$.
Three states are identified with $J^{PC} = 1^{--}$, having masses of 6524.51, 6555.97, and 6740.27 MeV, respectively. 
The two lower-mass states, at 6524.51 and 6555.97 MeV, lie above the $J/\psi\chi_{c0}$ threshold but below the $J/\psi\chi_{c1}$ threshold. 
Consequently, due to conservation laws, they are allowed to decay only into the $\eta_c h_c$ and $J/\psi\chi_{c0}$ channels.  
The highest $1^{--}$ state, at 6740.27 MeV, lies above the $J/\psi\chi_{c2}$ threshold and can decay into $\eta_c h_c$, $J/\psi\chi_{c0}$, $J/\psi\chi_{c1}$, and $J/\psi\chi_{c2}$ final states. 
Based on the properties of the final-state decay width, this state is more easily observed in the $J/\psi\chi_{c1}$ and $J/\psi\chi_{c2}$ channels compared to the $\eta_c h_c$ and $J/\psi\chi_{c0}$ channels.
In addition, there is a single $1^{-+}$ state with a mass of 6586.89 MeV, which can decay only into the $\eta_c\chi_{c1}$ channel.

\subsubsection{$[bb][\bar{b}\bar{b}]$ system}

For the tetraquark state $[bb][\bar{b}\bar{b}]$ with different quantum numbers, the mass spectra and corresponding meson-meson thresholds are exhibited in Fig.~\ref{fig:cccc bbbb cbcb}(b) and the decay channels are shown in Table~\ref{tab:cccc bbbb masses and decay}.

The quantum numbers of the $1S$-wave fully bottom tetraquark states are $J^{PC} = 0^{++}$, $1^{+-}$, and $2^{++}$. 
There are two $0^{++}$ states with masses of 18948.9 and 19066.9 MeV. The $1^{+-}$ and $2^{++}$ states have masses of 19006.9 and 19037.7 MeV, respectively. 
All these $S$-wave states lie above the $\Upsilon\Upsilon$ threshold. Both $0^{++}$ states are allowed to decay into the $\eta_b\eta_b$ and $\Upsilon\Upsilon$ channels. 
But this state is more likely to decay into the $\eta_b\eta_b$ channel due to its larger width. 
Correspondingly, these two states are more easily found in the $\Upsilon\Upsilon$ invariant mass spectrum.
The $1^{+-}$ state can decay only into the $\eta_b\Upsilon$ channel, while the $2^{++}$ state is allowed to decay only into the $\Upsilon\Upsilon$ channel.

The quantum numbers of the $P$-wave fully bottom tetraquark states are $J^{PC} = 0^{-+}$, $1^{--}$, and $1^{-+}$. 
Their masses all lie above the $\Upsilon\chi_{b2}$ threshold. 
Taking conservation laws into account, the $0^{-+}$ state can decay into the $\eta_b\chi_{b0}$ and $\Upsilon h_b$ channels, while the $1^{-+}$ state is allowed to decay into the $\eta_b\chi_{b1}$ and $\Upsilon h_b$ channels. 
The $1^{--}$ state can decay into multiple channels, including $\eta_b h_b$, $\Upsilon\chi_{b0}$, $\Upsilon\chi_{b1}$, and $\Upsilon\chi_{b2}$ channels.

\subsubsection{$[cb][\bar{c}\bar{b}]$ system}

For the tetraquark state $[cb][\bar{c}\bar{b}]$ with different quantum numbers, the mass spectra and corresponding meson-meson thresholds are exhibited in Fig.~\ref{fig:cccc bbbb cbcb}(c), and the decay channels are shown in Table~\ref{tab:cbcb masses and decay}.

Since the tetraquark $[cb][\bar{c}\bar{b}]$ system contains no identical quarks or antiquarks, it is not subject to the Pauli exclusion principle. 
As a result, there is no restriction on the construction of basis wave functions, and all possible spin-parity and charge-conjugation quantum numbers can be realized. 
This leads to a rich spectra of possible tetraquark states, including $0^{++}$, $1^{++}$, $1^{+-}$ and $2^{++}$ for $S$-wave, and $0^{--}$, $0^{-+}$, $1^{--}$ and $1^{-+}$ for $P$-wave.

There are four states with quantum numbers $0^{++}$, and their masses are 12477.8, 12548.6, 12633.4, and 12690.1 MeV, respectively. 
Among them, the two lighter states, 12477.8 and 12548.6 MeV, lie above the $\eta_c\eta_b$ threshold but below the $J/\psi\Upsilon$ threshold. 
Therefore, considering the conservation of angular momentum and $C$ parity, they can decay only into the $\eta_c\eta_b$ channel. 
In addition, the $[cb][\bar{c}\bar{b}]$ tetraquark states can decay not only into $c\bar{c}$ and $b\bar{b}$ mesons, but also into $c\bar{b}$ and $b\bar{c}$ mesons. 
However, since the masses of the two lighter states are below the $B_c\bar{B}_c$ threshold, decays into $c\bar{b}$ and $b\bar{c}$ mesons are kinematically forbidden. 
The two heavier states, with masses of 12633.4 and 12690.1 MeV, lie above the $J/\psi\Upsilon$ threshold and can, thus, decay into both $\eta_c\eta_b$ and $J/\psi\Upsilon$ channels. 
Moreover, since their masses are also above the $B_c\bar{B}_c$ threshold, decays into this channel are allowed. 
In particular, the state with mass 12690.1 MeV lies above the $B_c^*\bar{B}_c^*$ threshold and, therefore, can also decay into the $B_c^*\bar{B}_c^*$ channel. 
The two states are more likely to decay into mesons $\eta_c\eta_b$ due to their larger width. 
Similarly, due to the narrower decay width, 12633.4 MeV is more easily detected in the $J/\psi\Upsilon$ invariant mass spectrum, while 12690.1 MeV is more easily detected in the $B_c^*\bar{B}_c^*$ invariant mass spectrum.

For the quantum numbers $1^{++}$, there are two states with masses of 12582.8 and 12665.6 MeV, respectively. 
Both states lie above the $J/\psi\Upsilon$ threshold and are, therefore, allowed to decay into this channel. 
In addition, the heavier 12665.6 MeV lies above the thresholds for both $B_c\bar{B}_c^*$ and $B_c^*\bar{B}_c^*$, making decays into these channels kinematically allowed as well. 
But the width of $J/\psi\Upsilon$ channel is larger than $B_c\bar{B}_c^*$ and $B_c^*\bar{B}_c^*$; thus, the state of 12665.6 MeV is more likely to decay into $J/\psi\Upsilon$ channel.
Of course, the significantly narrower width indicates that this state is more likely to be observed in the $B_c^*\bar{B}_c^*$ invariant mass spectrum.
However, reconstructing the $B_c^*\bar{B}_c^*$ mesons is more challenging than the $J/\psi\Upsilon$ resonance. 
Therefore, unfortunately, the $[cb][\bar{c}\bar{b}]$ tetraquark state at 12665.6 MeV remains difficult to observe experimentally.

For the quantum numbers $1^{+-}$, there are four states with masses of 12547.4, 12590.8, 12617.6, and 12655.4 MeV, respectively. 
All of these states lie above the $\eta_c\Upsilon$ and $J/\psi\eta_b$ thresholds, allowing them to decay into both channels. 
Based on decay width, 12582.8 MeV is more likely to be found in the $J/\psi\eta_b$ invariant mass spectrum, while 12665.6 MeV is more likely to be found in the $B_c\bar{B}_c^*$ invariant mass spectrum.
Additionally, the two heavier states, with masses of 12617.6 and 12655.4 MeV, are above the $B_c\bar{B}_c^*$ threshold and can, therefore, also decay into this channel.
Both the four $1^{+-}$ states are more likely to decay into mesons $\eta_c\Upsilon$ because the width of $\eta_c\Upsilon$ is larger than that of other channels.
Similarly, calculations show that 12617.6 MeV is more readily observed in the $B_c\bar{B}_c^*$ invariant mass spectrum.

There are two states with the quantum numbers $2^{++}$, whose masses are 12649.1 and 12666.1 MeV, respectively. 
Both them lie above the $J/\psi\Upsilon$ threshold and are allowed to decay into this channel. 
Furthermore, the state with 12666.1 MeV lies above the threshold of $B_c^*\bar{B}_c^*$ and, therefore, can also decay into the $B_c^*\bar{B}_c^*$ channel. 
The 12666.1 MeV is more likely to decay into mesons $J/\psi\Upsilon$ due to the larger width.
Theoretically, the smaller width makes this state easier to observe in the final $B_c^*\bar{B}_c^*$ mesons.

There are two $0^{--}$ states, whose masses are 12955.1 and 13037.9 MeV. 
The mass of 13037.9 MeV lies above $J/\psi\chi_{b1}$, $\chi_{c1}\Upsilon$ and $B_c\bar{B}_{c0}$ thresholds; therefore, this state can naturally decay into these channels. 
Furthermore, due to its narrower width, this state is more easily detected in the $B_c\bar{B}_{c0}$ invariant mass spectrum.
The state with a mass of 12955.1 MeV has no available $S$-wave strong decay channels, Because its mass lies below $J/\psi\chi_{b1}$, $\chi_{c1}\Upsilon$, and $B_c\bar{B}_{c0}$ thresholds. 
However, it can decay into the $J/\psi\Upsilon$ channel via a $P$-wave transition.

For quantum numbers $0^{-+}$, there are four states with masses of 12919.7, 12963.1, 12989.9, and 13027.7 MeV, respectively.
Their masses lie all above the thresholds of $\eta_c\chi_{b0}$ and $\chi_{c0}\eta_b$; thus, they can naturally decay into these two channels. 
The heaviest state lies above the thresholds of the $J/\psi h_b$, $h_c\Upsilon$, and $B_c\bar{B}_{c0}$ channels and can, therefore, decay spontaneously into all three of them. 
The significantly narrower width suggests that 13027.7 MeV may form a peak in the $h_c\Upsilon$ invariant mass spectrum.
Besides, both the four states are more likely to decay into mesons $\chi_{c0}\eta_b$ due to the larger width.

There are eight $P$-wave tetraquark states with $J^{PC}=1^{--}$. 
Their masses and decay channels are exhibited in Table~\ref{tab:cbcb masses and decay}. 
Similarly, we can find some states with relatively narrow decay widths, which warrants closer attention in experiments. 
Of particular note is the fact that the tetraquark state with a mass of 13067.9 MeV is very close to the $B_c^*\bar{B}_{c1}$ threshold.
As a result, it can easily decay into the mesons $B_c^*\bar{B}_{c1}$.
And this state is expected to appear as a narrow resonance due to the proximity to the threshold.

For quantum numbers $1^{-+}$, there are four $[cb][\bar{c}\bar{b}]$ states with masses of 12949.7, 12993.1, 13019.9, and 13057.7 MeV.
All of these states lie above the $\eta_c\chi_{b1}$ and $\chi_{c1}\eta_b$ thresholds and can, therefore, naturally decay into these two channels. 
In addition, the state with a mass of 12993.1 MeV also lies above the $h_c\Upsilon$ threshold, making this decay channel accessible as well. 
In particular, the narrow decay width suggests that this state may form a peak in the $h_c\Upsilon$ invariant mass spectrum.
The two higher-mass states, 13019.9 and 13057.7 MeV, lie above even higher thresholds, including $\chi_{c1}\eta_b$ and $B_c\bar{B}_{c1}$. 
Furthermore, the heaviest state, with a mass of 13057.7 MeV, lies above the $B_c^*\bar{B}_{c0}$ threshold and can decay into this channel as well.
Of course, based on decay width analysis, 13019.9 and 13057.7 MeV are more likely to be found in the invariant mass spectra of $B_c\bar{B}_{c1}$ and $B_c^*\bar{B}_{c0}$, respectively.

\subsection{$[cc][\bar{c}\bar{b}]$, $[bb][\bar{b}\bar{c}]$, and $[cc][\bar{b}\bar{b}]$ systems}

For other fully heavy tetraquark $[cc][\bar{c}\bar{b}]$, $[bb][\bar{b}\bar{c}]$, and $[cc][\bar{b}\bar{b}]$ systems, $C$ parity cannot be well defined due to the violation of flavor symmetry in quark components.
As a result, the possible quantum numbers for these configurations are $J^P = 0^+$, $1^+$, $2^+$, $0^-$, and $1^-$. 
Our theoretical results are shown in Fig.~\ref{fig:cccb bbbc ccbb} and Tables~\hyperref[tab:cccb masses and decay]{\ref*{tab:cccb masses and decay}--\ref*{tab:ccbb masses and decay}}. 
The analytical method is similar to the previous tetraquark $[cc][\bar{c}\bar{c}]$, $[bb][\bar{b}\bar{b}]$, and $[cb][\bar{c}\bar{b}]$ systems. 
Therefore, we will not elaborate on it here.

\section{Summary \label{sec:summary}}

In this paper, we systematically study the mass spectra of the lower $S$- and $P$-wave fully heavy tetraquark states, as well as analyze their possible strong decay channels. 
Considering the basic features of the fully heavy tetraquark system, we adopted an improved Hamiltonian based on the one-gluon exchange process.
The Hamiltonian exhibits a diquark-antidiquark configuration, specifically including effective mass of constituent quark, spin-spin interaction, spin-orbit coupling, and pure orbital contribution.
The main model parameters required for our calculations are derived from fitting conventional mesons and baryons.
Other parameters $\kappa_{ij}$ corresponding to different color structures can be calculated through $SU(3)$ group algebra.

For the study of fully charmed tetraquark systems, our results do not support interpreting $X(6600)$, $X(6900)$, and $X(7200)$ as $1S$-wave fully charmed tetraquark states.
It is worth noting that we predict a $1S$-wave fully charmed tetraquark state $cc\bar{c}\bar{c}$ with quantum numbers $J^{PC} = 2^{++}$; its mass is 6260.89 MeV.
Meanwhile, we analyzed the decay channels of this tetraquark state and further confirmed that its characteristics are consistent with the observational results of the $X(6200)$ resonance found in the recent study of the LHCb dataset.
For the fully bottom tetraquark system, our calculated mass spectra are all above the $\Upsilon\Upsilon$ threshold and exceed the $bb\bar{b}\bar{b}$ mass reported by CMS Collaboration and A$_N$DY Collaboration. 
Nevertheless, LHCb Collaboration and other experiments have not observed any $bb\bar{b}\bar{b}$ resonant states thus far.  
In addition, we calculated masses of several other fully heavy tetraquark states and analyzed their strong decay channels. 
Finally, it is anticipated that some of the narrower fully heavy tetraquark states predicted in this work can be observed in future experiments.

\begin{acknowledgments}
	The authors thank Yanmei Xiao for helpful comments and discussions. This work was supported by the Scientific Research Foundation of Chengdu University of Technology under Grant No. 10912-KYQD2025-09557.
\end{acknowledgments}

\appendix 
\section*{APPENDIX: WAVE FUNCTIONS}

The basis wave functions of all fully heavy tetraquark states are given below.

(i) $Q_1Q_2\bar{Q}_3\bar{Q}_4$ with flavor=\{$cc\bar{c}\bar{c}$,$bb\bar{b}\bar{b}$,$cc\bar{b}\bar{b}$\}:
\begin{equation}
	\begin{gathered}
		|0^{+(+)}\rangle_1=|(Q_1Q_2)_0^6(\bar{Q}_3\bar{Q}_4)_0^{\bar{6}}\rangle_0,\\
		|0^{+(+)}\rangle_2=|(Q_1Q_2)_1^{\bar{3}}(\bar{Q}_3\bar{Q}_4)_1^3\rangle_0,
	\end{gathered}
\end{equation}

\begin{equation}
	|1^{+(-)}\rangle=|(Q_1Q_2)_1^{\bar{3}}(\bar{Q}_3\bar{Q}_4)_1^3\rangle_1,
\end{equation}

\begin{equation}
	|2^{+(+)}\rangle=|(Q_1Q_2)_1^{\bar{3}}(\bar{Q}_3\bar{Q}_4)_1^3\rangle_2,
\end{equation}

\begin{equation}
	|0^{-(+)}\rangle=|(Q_1Q_2)_1^{\bar{3}}(\bar{Q}_3\bar{Q}_4)_1^3\rangle_1,
\end{equation}

\begin{equation}
	\begin{gathered}
		|{1^{-(-)}}\rangle_1=|{(Q_1Q_2)_0^6(\bar{Q}_3\bar{Q}_4)_0^{\bar{6}}}\rangle_0,\\
		|{1^{-(-)}}\rangle_2=|{(Q_1Q_2)_1^{\bar{3}}(\bar{Q}_3\bar{Q}_4)_1^3}\rangle_0,\\
		|{1^{-(+)}}\rangle_3=|{(Q_1Q_2)_1^{\bar{3}}(\bar{Q}_3\bar{Q}_4)_1^3}\rangle_1,\\
		|{1^{-(-)}}\rangle_4=|{(Q_1Q_2)_1^{\bar{3}}(\bar{Q}_3\bar{Q}_4)_1^3}\rangle_2.
	\end{gathered}
\end{equation}

(ii) $Q_1Q_2\bar{Q}_3\bar{Q}_4$ with flavor=\{$cc\bar{c}\bar{b}$,$bb\bar{b}\bar{c}$\}:
\begin{equation}
	\begin{gathered}
		|{0^+}\rangle_1=|{(Q_1Q_2)_0^6(\bar{Q}_3\bar{Q}_4)_0^{\bar{6}}}\rangle_0,\\
		|{0^+}\rangle_2=|{(Q_1Q_2)_1^{\bar{3}}(\bar{Q}_3\bar{Q}_4)_1^3}\rangle_0,
	\end{gathered}
\end{equation}
\begin{equation}
	\begin{gathered}
		|1^+\rangle_1=|(Q_1Q_2)_0^6(\bar{Q}_3\bar{Q}_4)_1^{\bar{6}}\rangle_1,\\
		|1^+\rangle_2=|(Q_1Q_2)_1^{\bar{3}}(\bar{Q}_3\bar{Q}_4)_0^3\rangle_1,\\
		|1^+\rangle_3=|(Q_1Q_2)_1^{\bar{3}}(\bar{Q}_3\bar{Q}_4)_1^3\rangle_1,
	\end{gathered}
\end{equation}

\begin{equation}
	|2^+\rangle=|(Q_1Q_2)_1^{\bar{3}}(\bar{Q}_3\bar{Q}_4)_1^3\rangle_2,
\end{equation}

\begin{equation}
	\begin{gathered}
		|0^-\rangle_1=|(Q_1Q_2)_0^6(\bar{Q}_3\bar{Q}_4)_1^{\bar{6}}\rangle_1,\\
		|0^-\rangle_2=|(Q_1Q_2)_1^{\bar{3}}(\bar{Q}_3\bar{Q}_4)_0^3\rangle_1,\\
		|0^-\rangle_3=|(Q_1Q_2)_1^{\bar{3}}(\bar{Q}_3\bar{Q}_4)_1^3\rangle_1,
	\end{gathered}
\end{equation}

\begin{equation}
	\begin{gathered}
		|1^-\rangle_1=|(Q_1Q_2)_0^6(\bar{Q}_3\bar{Q}_4)_0^{\bar{6}}\rangle_0,\\
		|1^-\rangle_2=|(Q_1Q_2)_1^{\bar{3}}(\bar{Q}_3\bar{Q}_4)_1^3\rangle_0,\\
		|1^-\rangle_3=|(Q_1Q_2)_0^6(\bar{Q}_3\bar{Q}_4)_1^{\bar{6}}\rangle_1,\\
		|1^-\rangle_4=|(Q_1Q_2)_1^{\bar{3}}(\bar{Q}_3\bar{Q}_4)_0^3\rangle_1,\\
		|1^-\rangle_5=|(Q_1Q_2)_1^{\bar{3}}(\bar{Q}_3\bar{Q}_4)_1^3\rangle_1,\\
		|1^-\rangle_6=|(Q_1Q_2)_1^{\bar{3}}(\bar{Q}_3\bar{Q}_4)_1^3\rangle_2.
	\end{gathered}
\end{equation}

(iii) $cb\bar{c}\bar{b}$:
\begin{equation}
	\begin{gathered}
		|0^{++}\rangle_1=|(cb)_0^6(\bar{c}\bar{b})_0^{\bar{6}}\rangle_0,\\
		|0^{++}\rangle_2=|(cb)_0^{\bar{3}}(\bar{c}\bar{b})_0^3\rangle_0,\\
		|0^{++}\rangle_3=|(cb)_1^6(\bar{c}\bar{b})_1^{\bar{6}}\rangle_0,\\
		|0^{++}\rangle_4=|(cb)_1^{\bar{3}}(\bar{c}\bar{b})_1^3\rangle_0,
	\end{gathered}
\end{equation}

\begin{equation}
	\begin{gathered}
		|1^{++}\rangle_1=\frac{1}{\sqrt{2}}(|(cb)_1^6(\bar{c}\bar{b})_0^{\bar{6}}\rangle_1+|(cb)_0^6(\bar{c}\bar{b})_1^{\bar{6}}\rangle_1),\\
		|1^{++}\rangle_2=\frac{1}{\sqrt{2}}(|(cb)_1^{\bar{3}}(\bar{c}\bar{b})_0^3\rangle_1+|(cb)_0^{\bar{3}}(\bar{c}\bar{b})_1^3\rangle_1),
	\end{gathered}
\end{equation}

\begin{equation}
	\begin{gathered}
		|1^{+-}\rangle_1=\frac{1}{\sqrt{2}}(|(cb)_1^6(\bar{c}\bar{b})_0^{\bar{6}}\rangle_1-|(cb)_0^6(\bar{c}\bar{b})_1^{\bar{6}}\rangle_1),\\
		|1^{+-}\rangle_2=\frac{1}{\sqrt{2}}(|(cb)_1^{\bar{3}}(\bar{c}\bar{b})_0^3\rangle_1-|(cb)_0^{\bar{3}}(\bar{c}\bar{b})_1^3\rangle_1),\\
		|1^{+-}\rangle_3=|(cb)_1^6(\bar{c}\bar{b})_1^{\bar{6}}\rangle_1,\\
		|1^{+-}\rangle_4=|(cb)_1^{\bar{3}}(\bar{c}\bar{b})_1^3\rangle_1,
	\end{gathered}
\end{equation}

\begin{equation}
	\begin{gathered}
		|2^{++}\rangle_1=|(cb)_1^6(\bar{c}\bar{b})_1^{\bar{6}}\rangle_2,\\
		|2^{++}\rangle_2=|(cb)_1^{\bar{3}}(\bar{c}\bar{b})_1^3\rangle_2,
	\end{gathered}
\end{equation}

\begin{equation}
	\begin{gathered}
		|0^{--}\rangle_1=\frac{1}{\sqrt{2}}(|(cb)_1^6(\bar{c}\bar{b})_0^{\bar{6}}\rangle_1+|(cb)_0^6(\bar{c}\bar{b})_1^{\bar{6}}\rangle_1),\\
		|0^{--}\rangle_2=\frac{1}{\sqrt{2}}(|(cb)_1^{\bar{3}}(\bar{c}\bar{b})_0^3\rangle_1+|(cb)_0^{\bar{3}}(\bar{c}\bar{b})_1^3\rangle_1),
	\end{gathered}
\end{equation}

\begin{equation}
	\begin{gathered}
		|0^{-+}\rangle_1=\frac{1}{\sqrt{2}}(|(cb)_1^6(\bar{c}\bar{b})_0^{\bar{6}}\rangle_1-|(cb)_0^6(\bar{c}\bar{b})_1^{\bar{6}}\rangle_1),\\
		|0^{-+}\rangle_2=\frac{1}{\sqrt{2}}(|(cb)_1^{\bar{3}}(\bar{c}\bar{b})_0^3\rangle_1-|(cb)_0^{\bar{3}}(\bar{c}\bar{b})_1^3\rangle_1),\\
		|0^{-+}\rangle_3=|(cb)_1^6(\bar{c}\bar{b})_1^{\bar{6}}\rangle_1,\\
		|0^{-+}\rangle_4=|(cb)_1^{\bar{3}}(\bar{c}\bar{b})_1^3\rangle_1,
	\end{gathered}
\end{equation}

\begin{equation}
	\begin{gathered}
		|1^{--}\rangle_1=|(cb)_0^6(\bar{c}\bar{b})_0^{\bar{6}}\rangle_0,\\
		|1^{--}\rangle_2=|(cb)_0^{\bar{3}}(\bar{c}\bar{b})_0^3\rangle_0,\\
		|1^{--}\rangle_3=|(cb)_1^6(\bar{c}\bar{b})_1^{\bar{6}}\rangle_0,\\
		|1^{--}\rangle_4=|(cb)_1^{\bar{3}}(\bar{c}\bar{b})_1^3v_0,\\
		|1^{--}\rangle_5=\frac{1}{\sqrt{2}}(|(cb)_1^6(\bar{c}\bar{b})_0^{\bar{6}}\rangle_1+|(cb)_0^6(\bar{c}\bar{b})_1^{\bar{6}}\rangle_1),\\
		|1^{--}\rangle_6=\frac{1}{\sqrt{2}}(|(cb)_1^{\bar{3}}(\bar{c}\bar{b})_0^3\rangle_1+|(cb)_0^{\bar{3}}(\bar{c}\bar{b})_1^3\rangle_1),\\
		|1^{--}\rangle_7=|(cb)_1^6(\bar{c}\bar{b})_1^{\bar{6}}\rangle_2,\\
		|1^{--}\rangle_8=|(cb)_1^{\bar{3}}(\bar{c}\bar{b})_1^3\rangle_2,
	\end{gathered}
\end{equation}

\begin{equation}
	\begin{gathered}
		|1^{-+}\rangle_1=\frac{1}{\sqrt{2}}(|(cb)_1^6(\bar{c}\bar{b})_0^{\bar{6}}\rangle_1-|(cb)_0^6(\bar{c}\bar{b})_1^{\bar{6}}\rangle_1),\\
		|1^{-+}\rangle_2=\frac{1}{\sqrt{2}}(|(cb)_1^{\bar{3}}(\bar{c}\bar{b})_0^3\rangle_1-|(cb)_0^{\bar{3}}(\bar{c}\bar{b})_1^3\rangle_1),\\
		|1^{-+}\rangle_3=|(cb)_1^6(\bar{c}\bar{b})_1^{\bar{6}}\rangle_1,\\
		|1^{-+}\rangle_4=|(cb)_1^{\bar{3}}(\bar{c}\bar{b})_1^3\rangle_1.
	\end{gathered}
\end{equation}

\bibliography{ref}

\end{document}